# Theoretical design study of FWM in silicon nitride waveguides with integrated graphene oxide films

Yang Qu, Jiayang Wu, *Member, IEEE*, Yuning Zhang, Linnan Jia, Yao Liang, Baohua Jia, *Fellow, OSA*, and David J. Moss, *Fellow, IEEE, Fellow OSA*

*Abstract*—We theoretically investigate and optimize four-wave mixing (FWM) in silicon nitride (SiN) waveguides integrated with two-dimensional (2D) layered graphene oxide (GO) films. Based on extensive previous measurements of the material parameters of the GO films, we perform detailed analysis for the influence of device parameters including waveguide geometry, GO film thickness, length, and coating position on the FWM conversion efficiency (CE) and conversion bandwidth (CB). The influence of dispersion and photo-thermal changes in the GO films is also discussed. Owing to the strong mode overlap between the SiN waveguides and the highly nonlinear GO films, FWM in the hybrid waveguides can be significantly enhanced. We obtain good agreement with previous experimental results and show that by optimizing the device parameters to balance the trade-off between Kerr nonlinearity and loss, the FWM CE can be improved by as much as ~20.7 dB and the FWM CB can be increased by ~4.4 folds, relative to the uncoated waveguides. These results highlight the significantly enhanced FWM performance that can be achieved in SiN waveguides by integrating 2D layered GO films.

## I. Introduction

Four-wave mixing (FWM) is a parametric process that occurs in a nonlinear medium between incident light fields that interact to generate output light fields, potentially at new frequencies [1]. As a fundamental third-order ($\chi^{(3)}$) nonlinear optical process [2, 3], it has found wide applications in all-optical signal generation, amplification, and processing such as wavelength conversion [2, 4], optical frequency comb generation [5-7], optical sampling [8, 9], quantum entanglement [10-13], and many others [14-17].

Implementing FWM devices in integrated platforms would provide significant advantages in terms of compact footprint, high stability, high scalability, and mass-producibility [18-20]. This, along with the ultrafast response time of FWM (typically on the order of $10^{-15}$ s [21]), can enable devices with ultrahigh processing speeds and ultralarge operation bandwidths. Despite the fact that silicon has been a leading integrated platform for many FWM devices [20, 22], its FWM performance is well known to be significantly limited by strong two-photon absorption (TPA) in the telecommunications band. Other complementary metal-oxide-semiconductor (CMOS) compatible platforms such as silicon nitride (SiN) and doped silica [19, 23] have negligible TPA, although their Kerr nonlinearity is only moderately higher than silica glass, thus representing a limitation in terms of FWM efficiency.

To improve the performance of the state-of-the-art integrated FWM devices, two-dimensional (2D) materials with an ultrahigh optical nonlinearity have been integrated onto chips to implement hybrid devices. Recently, hybrid integrated photonics devices incorporating graphene, graphene oxide (GO), black phosphorus, and transition metal dichalcogenides (TMDCs) with high FWM performance have been reported [24-28].

Benefiting from its ease of preparation as well as the tunability of its material properties, GO has been recognized as a promising 2D material [29, 30]. Previously, GO films have been reported with a giant Kerr nonlinearity ($n_2$) of about 5 orders of magnitude higher than that of SiN [31, 32]. The relatively large bandgap of GO (> 2 eV [29, 33]) results in a material absorption that is over 2 orders of magnitude lower than graphene as well as negligible TPA in the telecommunications band [26]. Low linear and nonlinear loss are highly desirable for nonlinear applications such as FWM. Based on these material properties, enhanced FWM in GO-coated SiN and doped silica waveguides has been demonstrated [26, 34]. Another important advantage of GO is the ability to accurately control the film thickness, size and placement on integrated devices by using a large-area, transfer-free, layer-by-layer coating method together with standard lithography and lift-off processes [26, 35, 36]. This overcomes the limitations of mechanical layer transfer approaches that have been widely used for other 2D materials such as graphene and TMDCs [37] and enables cost-effective, large-scale, and highly precise integration of 2D layered GO films on a chip.

Previously [34], we experimentally demonstrated an enhancement of up to ~9.1 dB in the FWM conversion efficiency (CE) of SiN waveguides integrated with layered GO films. We obtained extensive measurements of the loss and Kerr nonlinearity ($n_2$) of the GO films that varied with film thickess and optical power. Here, we use these results to theoretically optimize the FWM performance of GO-coated

This work was supported by the Australian Research Council Discovery Projects Programs (No. DP150102972 and DP190103186), the Swinburne ECR-SUPRA program, the Industrial Transformation Training Centers scheme (Grant No. IC180100005), and the Beijing Natural Science Foundation (No. Z180007). *(Corresponding author: Jiayang Wu, Baohua Jia, and David J. Moss)*

Yang Qu, Jiayang Wu, Yuning Zhang, Linnan Jia and D. J. Moss are with Optical Sciences Centre, Swinburne University of Technology, Hawthorn, VIC 3122, Australia. (e-mail: jiayangwu@swin.edu.au; ljia@swin.edu.au and dmoss@swin.edu.au).

Yao Liang and Baohua Jia are with Centre for Translational Atomaterials, Swinburne University of Technology, Hawthorn, VIC 3122, Australia. (e-mail: bjia@swin.edu.au).

SiN waveguides. A detailed analysis of the influence of the device parameters, including waveguide geometry and GO film thickness, length, and coating position, on the FWM CE and conversion bandwidth (CB) is performed. The influence of dispersion and any photo-thermal changes in the GO films is also considered. We achieve good agreement with the previous experimental results and show that the enhancement in the FWM CE can be increased to as much as ~20.7 dB and the FWM CB can be increased by ~4.4 folds, by optimizing the device parameters to balance the trade-off between Kerr nonlinearity and loss. These results highlight the strong potential for high FWM performance of SiN waveguides integrated with 2D layered GO films.

## II. DEVICE STRUCTURE

Fig. 1(a) shows a schematic of a SiN waveguide coated with a patterned GO film, along with a schematic showing the atomic structure of GO with different oxygen functional groups (OFGs). In contrast to graphene that has a metallic behavior with zero bandgap [38], GO is a dielectric with a typical bandgap > 2 eV [29, 33] – much larger than what is expected for efficient TPA in the telecom band. SiN waveguides have been fabricated via low-pressure chemical vapor deposition (LPCVD) followed by lithography and dry etching processes, all of which are compatible with CMOS fabrication [39, 40]. To enable film coating on the top surface of the SiN waveguides, the silica upper cladding is removed, typically via chemical-mechanical polishing (CMP). The coating of GO films can be achieved via a solution-based method, which yields layer-by-layer film deposition [26, 35, 36]. Compared with the complex layer transfer processes widely employed in coating other 2D materials such as graphene and TMDCs [24, 41, 42], this approach enables transfer-free coating of highly uniform films over large areas, with highly scalable fabrication processes as well as precise control of the number of layers (i.e., film thickness). Film patterning can be achieved using standard lithography and lift-off processes to achieve precise control of the film length and position [36] in order to optimize the FWM performance of the hybrid waveguides.

Fig. 1(b) shows the schematic cross section of the hybrid waveguide in Fig. 1(a). The corresponding transverse electric (TE) mode profile is shown in Fig. 1(c). Nonlinear optical processes such as FWM in the highly nonlinear GO film can be excited by the evanescent field of the waveguide mode, forming the basis for enhancing nonlinear optical performance of the hybrid waveguides.

Table I shows the definition of the parameters that we use to investigate FWM in the GO-coated SiN waveguides, including waveguide dimensions ($W$ and $H$), GO film parameters ($N$, $L$, and $L_0$), and continuous-wave (CW) optical powers ($P_p$ and $P_s$). In the following sections, we first investigate the influence of mode overlap with the films, including the waveguide geometry ($W$ and $H$) and film thickness ($N$), on the loss and nonlinear parameter ($\gamma$) of the hybrid waveguides. Next, by properly balancing the trade-off between loss and nonlinearity, we optimize the FWM CE in the hybrid waveguides for different waveguide geometries ($W$, $H$) and GO film parameters ($N$, $L$, and $L_0$). Finally, we discuss the influence of the dispersion and any photo-thermal changes in the GO films in our model, as well as the influence of the film length on FWM CB and idler power evolution along the hybrid waveguides.

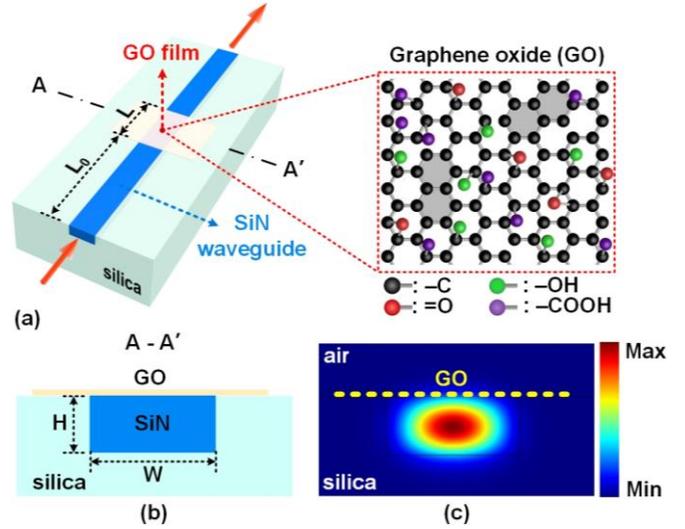

Fig. 1. (a) Schematic illustration of a SiN waveguide coated with 10 layers of GO. Inset shows schematics of atomic structure of GO. (b) Schematic illustration of the cross section of the hybrid waveguide in (a). (c) TE mode profile corresponding to (b). The definitions of $L_0$, $L$, $W$, and $H$ are given in Table I.

TABLE I. DEFINITIONS OF PARAMETERS OF WAVEGUIDES DIMENSION, GO FILM, AND CW LASER.

| Waveguide dimension | Height | Width |
|---|---|---|
| Parameters | $H$ | $W$ |
| GO film | GO layer number | Uncoated length before GO segment | Coating length |
| Parameters | $N$ | $L_0$ | $L$ |
| CW laser | Coupled pump power | Coupled signal power |
| Parameters | $P_p$ | $P_s$ |

## III. LOSS & NONLINEAR PARAMETER

Mode overlap is a key factor that influences both loss and the nonlinear parameter ($\gamma$) of GO-coated SiN waveguides. In this section, we investigate this using commercial mode solving softwares along with experimentally measured material parameters at 1550 nm.

Fig. 2(a) shows the extinction coefficient $k$ of GO versus layer number ($N$) for different input CW powers, which was obtained from our previous measurements [26, 36]. The $k$ of GO is over two orders of magnitudes lower than that of graphene [43, 44], reflecting its potential for high-performance nonlinear photonic devices. The increase of *k* with *N* can be attributed to scattering loss stemming from film unevenness and imperfect contact between adjacent layers [26, 36]. The small but observable increase in *k* with input CW power is mainly induced by power-dependent photo-thermal changes in the GO films [36, 45], which are not permanent and have a much slower time response than the nearly instantaneous nonlinear optical processes governing FWM and TPA [34, 46]. In contrast to *k*, the refractive index *n* of GO showed negligible variation with *N* and the input power [34], and so was taken as a constant (1.97) in our simulations.

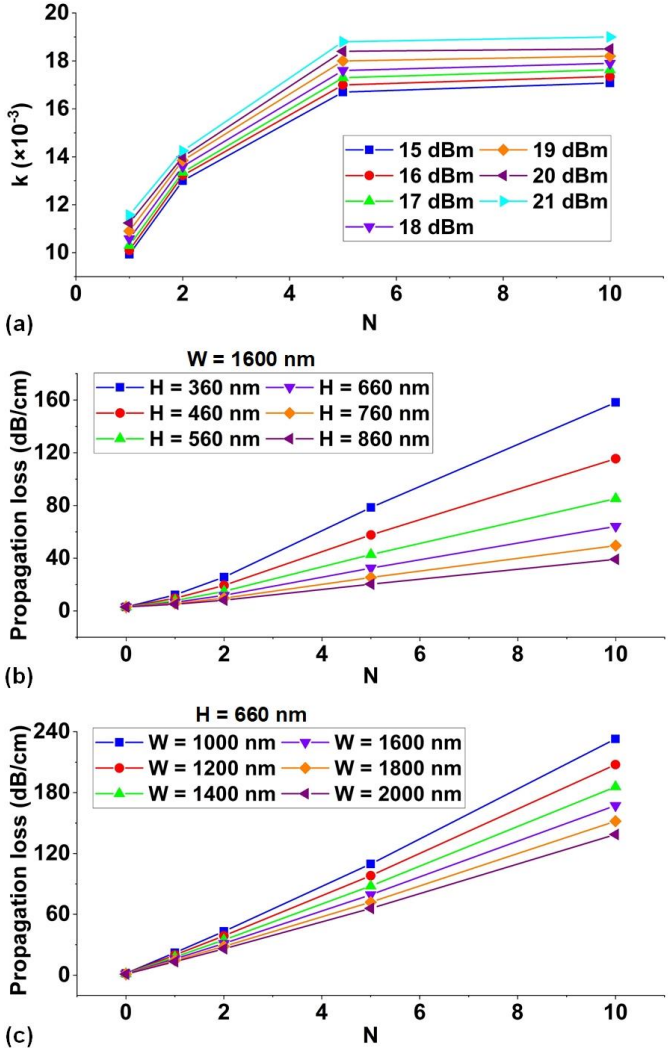

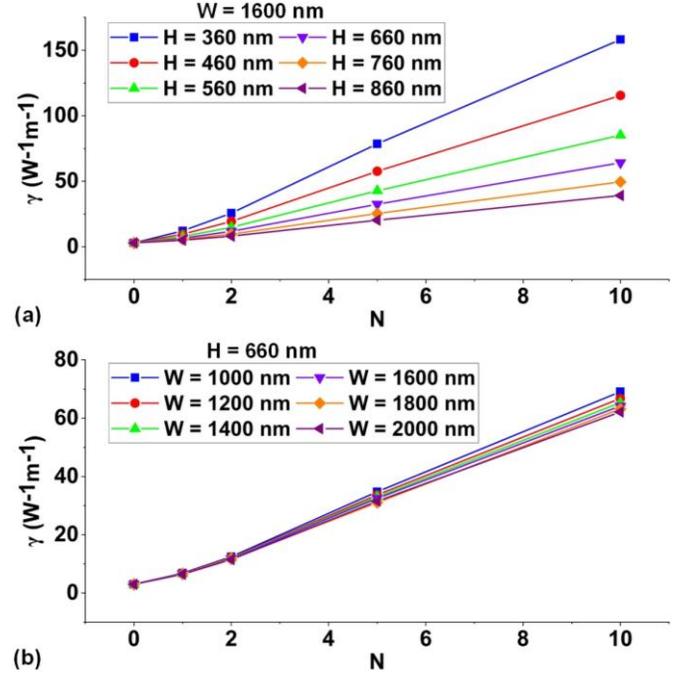

Fig. 2. (a) Extinction coefficient *k* of GO versus *N* under various input CW powers. (b) Propagation loss versus *N* for GO-coated SiN waveguides with various *H* when *W* = 1600 nm. (c) Propagation loss versus *N* for GO-coated SiN waveguides with various *W* when *H* = 660 nm. The points at *N* = 0 correspond to the results for bare SiN waveguides.

Fig. 3. Nonlinear parameter *γ* versus *N* for GO-coated SiN waveguides with (a) various *H* when *W* = 1600 nm (b) various *W* when *H* = 660 nm. The points at *N* = 0 correspond to the results for bare SiN waveguides.

Figs. 2(b) and (c) show the propagation loss of the hybrid waveguides versus GO layer number (*N*), first for different waveguide heights (*H*) with a fixed width (*W*) and then for different widths (*W*) with a fixed height (*H*), calculated using Lumerical FDTD Mode Solutions. We used the experimentally measured values for the linear optical properties of both the GO films and the SiN waveguides, including the above *n*, *k* of GO as well as previously fabricated SiN devices for the refractive index *n* of SiN (1.99) and propagation loss of the bare waveguides (3.0 dB/cm) [34]. Note that the propagation loss is higher than buried SiN waveguides (0.6 dB/cm) due to residual roughness induced by CMP, exacerbated by the higher SiN – air refractive index difference. According to our previous measurements [26, 36], the film thickness is assumed to be proportional to *N*, with a thickness of 2 nm for each layer. TE polarization was chosen because the in-plane (TE polarized) interaction between the evanescent field and the GO film is much stronger than the out-of-plane (TM polarized) interaction due to the large optical anisotropy of 2D materials [26, 37].

In Fig. 2(b), as the height of the waveguide *H* increases, the mode overlap with the films decreases, resulting in a decrease in propagation loss. At the same time, the propagation loss increases with GO layer number *N*, reflecting an increase in mode overlap for the hybrid waveguides having thicker films. Similar to the trends in Fig. 2(b), the propagation loss in Fig. 2(c) decreases with the waveguide width *W* and increases with GO layer number *N*, further confirming that the propagation loss increases with mode overlap with the GO films.

Figs. 3(a) and (b) show the nonlinear paremeter of the hybrid waveguides *γ* versus GO layer number (*N*) for different waveguide geometries (*H*, *W*), calculated based on methods described elsewhere [26, 46, 47]:

$$\gamma = \frac{2\pi}{\lambda} \frac{\iint_D n_0^2(x,y) n_2(x,y) S_z^2 \, dxdy}{\left[\iint_D n_0(x,y) S_z \, dxdy\right]^2} \quad (1)$$

where *λ* is the pump wavelength (i.e., 1550 nm), *D* is the integral domain of the optical fields over the material regions, $S_z$ is the time-averaged Poynting vector calculated using COMSOL Multiphysics, $n_0(x, y)$ and $n_2(x, y)$ are the linear and nonlinear refractive index profiles over the waveguide cross section, respectively. In Eq. (1), *γ* is an effective nonlinear parameter weighted not only by $n_2$ $(x, y)$ but also by $n_0$ $(x, y)$ in the different material regions. Note we used the Kerr nonlinearity $n_2$ $(x, y)$ instead of the more general third-order nonlinearity ($\chi^{(3)}$) because the FWM frequencies (pump, signal, idler) in our analysis are close enough together (with a pump-signal wavelength detuning < 40 nm) compared with any dispersion in $n_2$ [26, 48]. The values of $n_2$ for silica and SiN used in our calculations were $2.60 \times 10^{-20}$ m$^2$/W [19] and $2.61 \times 10^{-19}$ m$^2$/W [34], respectively. The layer-dependent $n_2$ for GO was obtained from our previous experiments [34], which varied from $1.40 \times 10^{-14}$ m$^2$/W for $N$ = 1 to $1.34 \times 10^{-14}$ m$^2$/W for $N$ = 10 and were mainly induced by an increase in inhomogeneous defects and imperfect contact with film thickness. Since the experimentally measured $n_2$ of GO showed very little variation with optical power, we neglected this in our calculations.

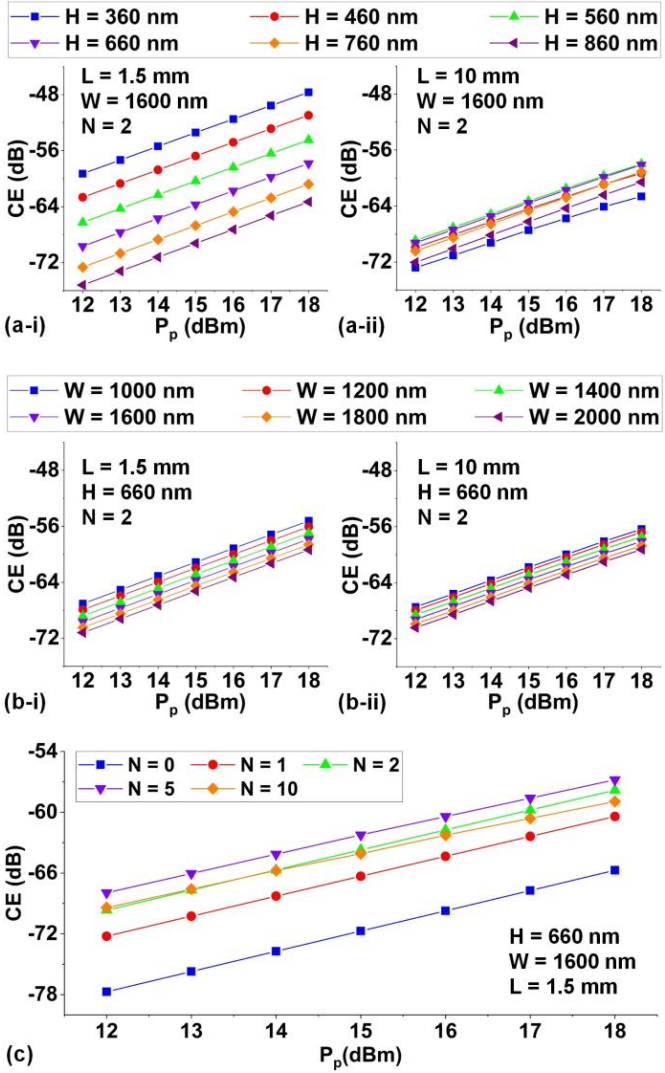

Fig. 4. FWM CE versus $P_p$ for GO-coated SiN waveguides with various (a) $H$, (b) $W$, and (c) $N$. In (a) and (b), $N$ = 2, (i) and (ii) show the results for $L$ = 1.5 mm and $L$ = 10 mm, respectively. In (a) and (c), $W$ = 1600 nm. In (b) and (c), $H$ = 660 nm. In (a) − (c), $L_0$ = 0.

In Figs. 3(a) and (b), we see that $\gamma$ increases with film thickness $N$ and decreases with waveguide height $H$ and width $W$, following the trends also seen with the propagation loss in Figs. 2(b) and (c). This indicates that increasing the mode overlap with GO increases both the Kerr nonlinearity and linear loss − a trade-off that needs to be properly balanced in order to optimize the nonlinear optical performance such as FWM.

## IV. CONVERSION EFFICIENCY

In this section, we investigate the influence of waveguide geometry ($W$, $H$) and GO film parameters ($N$, $L$, and $L_0$) on the FWM CE using the loss and $\gamma$ from section III. Considering only the idler at the shorter wavelength, and neglecting any depletion in the pump power due to idler generation, the FWM process in the GO-coated SiN waveguides can be modeled as [49, 50]

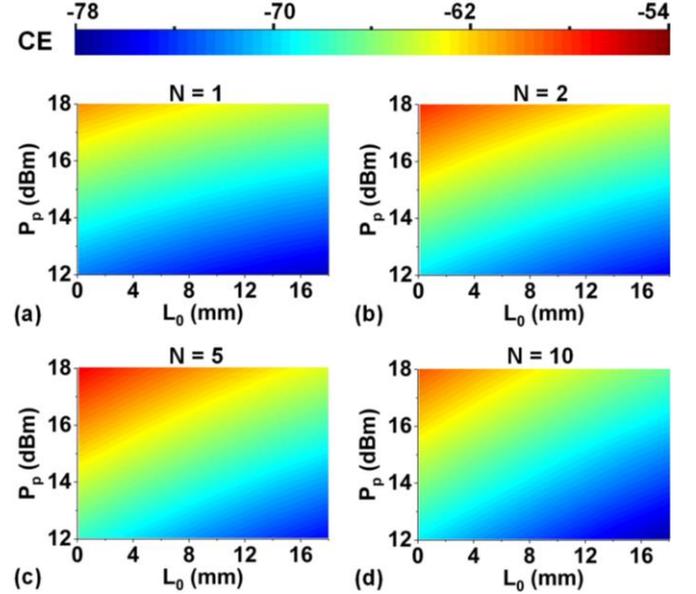

Fig. 5. FWM CE of GO-coated SiN waveguides versus $L_0$ and $P_p$ when (a) $N$ = 1, (b) $N$ = 2, (c) $N$ = 5, and (d) $N$ = 10. In (a) − (d), $L$ = 1.5 mm, $H$ = 660 nm, and $W$ = 1600 nm.

$$\frac{dA_p(z)}{dz} = -\frac{\alpha_p}{2}A_p(z) + j\gamma_p\left[|A_p(z)|^2 + 2|A_s(z)|^2 + 2|A_i(z)|^2\right]A_p(z)$$
$$+ j2\gamma_p A_p^*(z)A_s(z)A_i(z)exp(j\Delta\beta z) \quad (2)$$

$$\frac{dA_s(z)}{dz} = -\frac{\alpha_s}{2}A_s(z) + j\gamma_s\left[|A_s(z)|^2 + 2|A_p(z)|^2 + 2|A_i(z)|^2\right]A_s(z)$$
$$+ j\gamma_s A_i^*(z)A_p^2(z)exp(-j\Delta\beta z) \quad (3)$$

$$\frac{dA_i(z)}{dz} = -\frac{\alpha_i}{2}A_i(z) + j\gamma_i\left[|A_i(z)|^2 + 2|A_p(z)|^2 + 2|A_s(z)|^2\right]A_i(z)$$
$$+ j\gamma_i A_s^*(z)A_p^2(z)exp(-j\Delta\beta z) \quad (4)$$

where $A_{p,s,i}$ are the pump, signal, and idler amplitudes along the light propagation direction (i.e., $z$ axis), $\alpha_{p,s,i}$ are the linear losses in Fig. 2, $\gamma_{p,s,i}$ are the waveguide nonlinear parameters in Fig. 3, and $\Delta\beta = \beta_s + \beta_i - 2\beta_p$ is the linear phase mismatch, with $\beta_{p,s,i}$ denoting the propagation constants. In our case the wavelength detuning range was small ($\leq$ 40 nm), and so the linear loss and nonlinear parameter can be assumed to be wavelength independent, i.e., $\alpha_p = \alpha_s = \alpha_i = \alpha$, $\gamma_p = \gamma_s = \gamma_i = \gamma$. By numerically solving Eqs. (2) – (4), the FWM CE was calculated according to

$$CE \text{ (dB)} = 10 \times \log_{10}[|A_i(L_t)|^2/|A_s(0)|^2] \quad (5)$$

where $L_t$ is the total length of the SiN waveguide, taken as 20 mm in our analysis. The hybrid waveguides were divided into bare SiN (without GO films) and hybrid (with GO films) segments with different $\alpha$, $\gamma$, and $\beta_{p,s,i}$. Eqs. (2) – (4) were solved for each segment, and the output from the previous segment was used as the input for the subsequent segment.

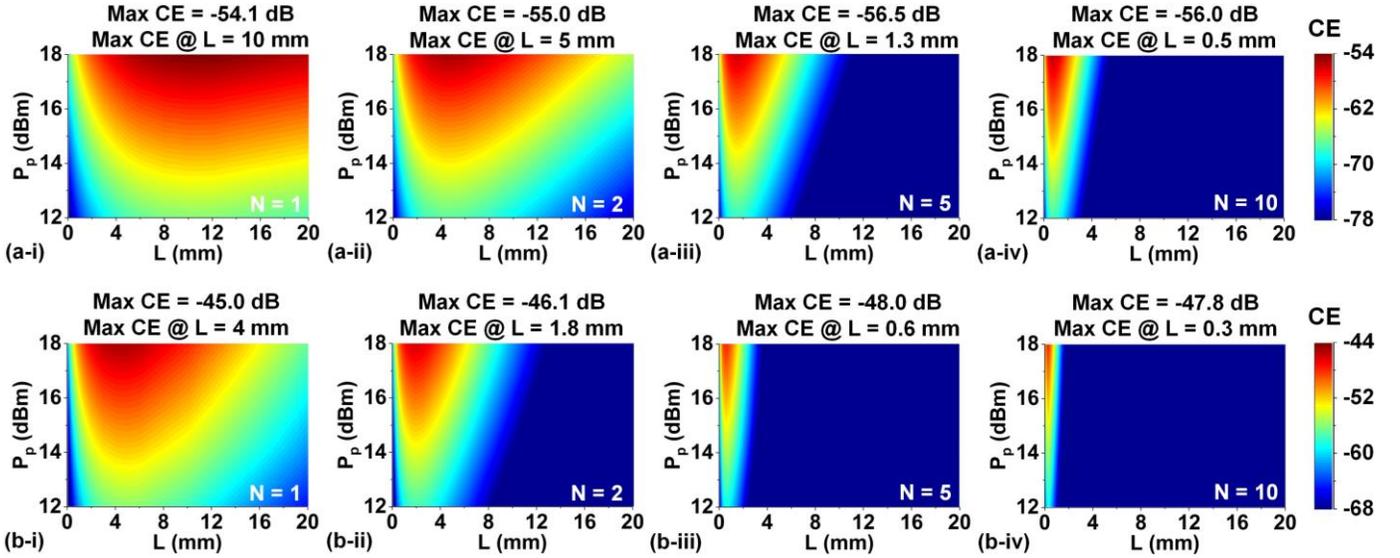

Fig. 6. FWM CE of GO-coated SiN waveguides versus $L$ and $P_p$ when (a) $W$ = 1600 nm, $H$ = 660 nm and (b) $W$ = 1000 nm, $H$ = 360 nm. (i) – (iv) show the results for $N$ = 1, 2, 5, and 10, respectively. In (a) and (b), $L_0$ = 0 mm.

In Figs. 4 (a) – (c), we compared the FWM CE of the hybrid waveguides while varying three parameters including the waveguide height and width ($H$, $W$) and the GO film thickness ($N$). In each figure, we varied only one parameter, keeping the other two constant. Fig. 4(a) compares the CE of the hybrid waveguides with the same width $W$ but different height $H$. The CE decreases with $H$ for short GO film lengths (i.e., $L$ = 1.5 mm in Fig. 4(a-i)), reflecting that the increase in $\gamma$ dominates the trade-off between the Kerr nonlinearity and loss. Whereas for long GO film lengths (i.e., $L$ = 10 mm in Fig. 4(a-i)), the CE varies non-monotonically with $H$, resulting in a maximum CE at an intermediate height $H$ = 560 mm. This is because the impact of loss becomes more prominent as $L$ increases. Fig. 4(b) compares the CE of the hybrid waveguides for a fixed height $H$ but different widths $W$. The CE decreases with $W$ monotonically for both $L$ = 1.5 mm and $L$ = 10 mm, while the decrease becomes more gradual for $L$ = 10 mm. This reflects the fact that the CE is more sensitive to changes in $H$ rather than $W$, which is consistent with what is seen for the linear loss in Fig. 2 and $\gamma$ in Fig. 3. Fig. 4(c) compares the CE of the hybrid waveguides for different film thicknesses $N$. The maximum CE is also achieved at an intermediate $N$ = 5 as a result of the trade-off between $\gamma$ and loss. In contrast to Figs. 4(a) and (b), the significant change in CE with $N$ results in an obvious trade-off between $\gamma$ and loss for a short GO film length of $L$ = 1.5 mm.

Figs. 5(a) – (d) show the CE of the hybrid waveguides versus coating position $L_0$ and pump power $P_p$ with $N$ = 1, 2, 5, and 10, respectively. For comparison, the other device parameters are kept the same as $L$ = 1.5 mm, $H$ = 660 nm, and $W$ = 1600 nm. In each figure, the CE increases with $P_p$ and decreases with $L_0$. The former reflects the increase in nonlinear efficiency with pump power while the latter results from a decrease in pump power in the GO-coated segment that dominates the FWM process in the hybrid waveguides. In contrast, for other waveguides than those considered here, such as for doped silica waveguides where the propagation loss of the bare waveguides is much lower, at 0.24 dB/cm [26], the effect of varying the coating position $L_0$ would be much smaller.

Fig. 6(a) shows the CE of the hybrid waveguides versus film length $L$ and pump power $P_p$ with (i) $N$ = 1, (ii) $N$ = 2, (iii) $N$ = 5, and (iv) $N$ = 10. For comparison, the other device parameters are kept the same as $L_0$ = 0 mm, $H$ = 660 nm, and $W$ = 1600 nm. In each figure, the CE increases with $P_p$, consistent with the trends seen in Fig. 5. The CE first increases with film length $L$ and then decreases, achieving maximum values at intermediate $L$ = 10 mm, 5 mm, 1.3 mm, and 0.5 mm, respectively. This reflects the fact that the enhancement in the Kerr nonlinearity dominates for short film lengths while the loss dominates for long film lengths. As $N$ increases, the maximum CE shifts towards shorter lengths, reflecting the fact that the influence of loss increase becomes more significant for the hybrid waveguides having thicker GO films. The CE of the hybrid waveguide is -56.6 dB when $N$ = 5, $L$ = 1.5 mm, and $P_p$ = 18 dBm, which is 9.1 dB higher than comparable devices without GO films, showing a good agreement with our previous experimental results [34]. The maximum CE enhancement of 11.6 dB is achieved when $N$ = 1, $L$ = 10 mm, and $P_p$ = 18 dBm, reflecting that there is still room for improvement in the experimental results by optimizing the GO film length and thickness.

Fig. 6(b) shows the corresponding results for hybrid waveguides with $H$ = 360 nm and $W$ = 1000 nm that have an enhanced mode overlap with GO films. The maximum CE of 45.0 dB is achieved when $N$ = 1 and $L$ = 4 mm, which is 20.7 dB higher than the bare waveguides and corresponds to 9.1 dB further improvement over the the maximum CE obtained in Fig. 6(a), reflecting the potential to improve the CE by optimizing the waveguide geometry.

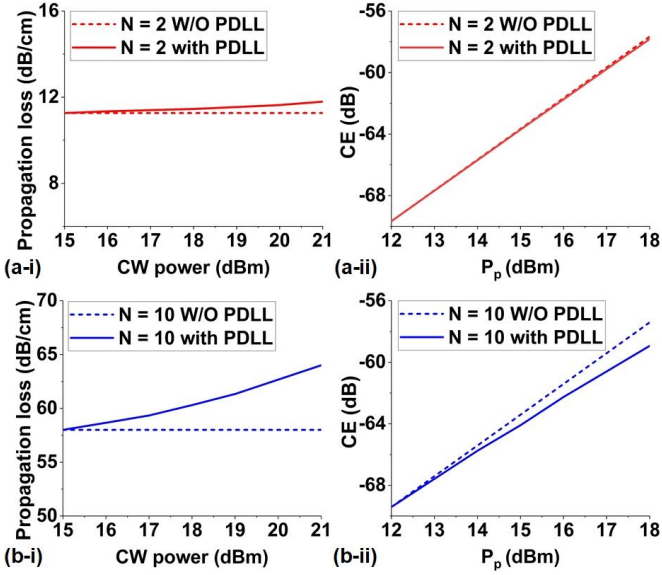

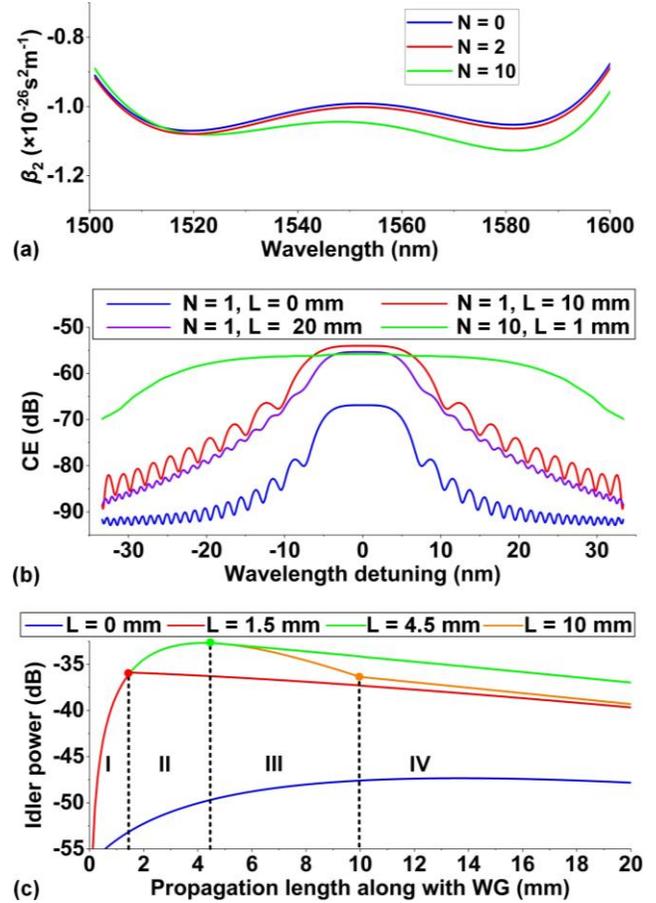

Fig. 7. (a) − (b) Performance comparison of GO-coated SiN waveguides with and without considering GO-induced power dependent linear loss (PDLL) when $N = 2$ and $N = 10$, respectively. (i) shows propagation loss versus input CW power and (ii) shows CE versus $P_p$. In (a) − (b), $L_0 = 0$ mm, $L = 1.5$ mm, $W = 1600$ nm, and $H = 660$ nm.

## V. DISCUSSION

In this section, we investigate the influence of the GO coating length, film dispersion, and power dependent loss induced by any photo-thermal changes on the FWM performance, focusing on the CB and idler power evolution along the hybrid waveguides.

Fig. 7 shows the influence on the CE of the power-dependent waveguide propagation loss induced by photo-thermal changes in the GO films. Since this has a much slower time response than the nonlinear optical processes of FWM and TPA [34, 46], we accounted for it by adjusting the loss (i.e., $\alpha$) of the hybrid waveguides in Eqs. (2) − (4) as a power dependent linear loss (PDLL). In Fig. 7(a), the solid curves show the results for $N = 2$ when considering the PDLL that slightly increases with input CW power, whereas the dashed curves represent results that were calculated using the $k$ value of GO at low powers (i.e., the results for 15 dBm in Fig. 2) without considering the PDLL. The FWM CE slightly decreases when including PDLL, with the difference becoming more significant at higher powers, consistent with the increase of PDLL with CW power. Fig. 7(b) shows the corresponding results when $N = 10$, where the difference is more obvious compared to the results for $N = 2$, reflecting that there is a more significant influence of the PDLL for the hybrid waveguides with thicker films.

Fig. 8(a) shows the group-velocity dispersion $\beta_2$ for the bare ($N = 0$) and coated SiN waveguides with 2 and 10 layers of GO ($N = 2, 10$). As compared with the bare waveguide, the hybrid waveguides display slightly enhanced anomalous dispersion, which becomes more significant as $N$ increases. This could result in better phase matching for FWM and add to the strong Kerr nonlinearity of the GO films.

Fig. 8(b) compares the CB for the bare ($N = 0$, $L = 0$ mm) versus hybrid waveguides with both uniformly coated ($N = 1$, $L = 20$ mm) and patterned ($N = 1$, $L = 10$ mm and $N = 10$, $L = 1$ mm) films. The maximum CE is achieved at $N = 1$, $L = 10$ mm

Fig. 8. (a) Group-velocity dispersion for bare and GO-coated SiN waveguides with 2 and 10 layers of GO. (b) CE versus wavelength detuning for bare SiN waveguides ($N = 0$, $L = 0$ mm) and hybrid waveguides with both uniformly coated ($N = 1$, $L = 20$ mm) and patterned ($N = 1$, $L = 10$ mm and $N = 10$, $L = 1$ mm) GO films. The pump wavelength is fixed at 1550 nm and the signal wavelength was detuned around 1550 nm. (c) Evolutions of idler power along the hybrid waveguides with different $L$ when $N = 2$. In (a) − (c), $L_0 = 0$ mm, $W = 1600$ nm, and $H = 660$ nm

due to a better trade-off between $\gamma$ and loss. In theory, the CB can be approximated by [2]:

$$CB \approx \left[\frac{4\pi}{\beta_2 L}\right]^{\frac{1}{2}} \quad (6)$$

which is inversely proportional to the square root of the product of $\beta_2$ and $L$. In Fig. 8(b), the maximum CB is achieved at $N = 10$, $L = 1$ mm, which is 4.4 times and 4.0 times that of the bare waveguides and the uniformly coated device, respectively. As reflected by Eq. (6), this mainly results from the shorter film length, with a small contribution from the slightly enhanced anomalous dispersion in Fig. 8(a). The broadened CB could also reflect the fact that most of the FWM occurs in the coated segment, where the efficiency is enhanced due to the better phase matching in a shorter coherence length.

Fig. 8(c) shows the evolution of the FWM idler power along the bare ($L = 0$) and hybrid waveguides with different film length $L = 1.5$ mm, 4.5 mm, and 10 mm. The SiN waveguide was divided into four regions I – IV according to different $L$. It can be seen that the idler power initially rises and then drops after reaching a peak. In the GO coated segment, the idler power reaches a maximum at $L = 4.5$ mm due to the trade-off between $\gamma$ and loss, whereas in the uncoated segment the idler power decreases continuously because of the propagation loss and relatively low Kerr nonlinearity of the bare SiN waveguides. This further confirms that the enhancement in the Kerr nonlinearity dominates for short film lengths whereas the loss dominates for long GO film lengths. It also provides a criteria for choosing the optimized GO film length in order to maximize the output idler power and hence the CE, and has implications for more advanced photonic integrated circuits [51, 52].

## VI. Conclusion

In summary, we theoretically optimize FWM in SiN waveguides integrated with 2D layered GO films. A detailed analysis of the influence of device parameters including waveguide geometry, GO film thickness, length, and position is performed, together with consideration of the influence of GO's dispersion and photo-thermal changes on the FWM performance. By optimizing the device parameters to balance the trade-off between Kerr nonlinearity and loss, up to ~20.7 dB improvement in FWM efficiency and ~4.4 times increase in FWM bandwidth are achieved. These results highlight the significantly enhanced FWM performance that can be achieved in SiN waveguides by integrating them with 2D layered GO films and serve as a roadmap to optimize the overall device performance.